\documentclass[a4paper]{article}
\usepackage{amscd,amsmath,amssymb,enumerate}
\usepackage{xcolor}

\newcommand{\p}[1]{(\ref{#1})}

\newcommand{\cN}{{\cal N}}

\newcommand{\cS}{{\cal S}}

\newcommand{\bD}{{\overline D}{}}

\newcommand{\bQ}{{\overline Q}{}}
\newcommand{\bS}{{\overline S}{}}

\newcommand{\bxi}{{\bar\xi}}

\newcommand{\bphi}{{\bar\phi}}
\newcommand{\bpsi}{{\bar\psi}{}}

\newcommand{\bv}{{\bar v}}

\newcommand{\be}{\begin{equation}}
\newcommand{\ee}{\end{equation}}
\newcommand{\bea}{\begin{eqnarray}}
\newcommand{\eea}{\end{eqnarray}}

\newcommand{\ba}{\begin{array}} \newcommand{\ea}{\end{array}}

\def\im{{\rm i\,}}
\def\sfrac#1#2{{\textstyle\frac{#1}{#2}}}

\newcommand{\nn}{\nonumber}

\begin{document}
\begin{flushright}
\end{flushright}\vspace{1cm}
\begin{center}
{\Large\bf  Generalized Schwarzians}
\end{center}
\vspace{1cm}

\begin{center}
{\Large\bf Nikolay Kozyrev and  Sergey Krivonos
}
\end{center}

\vspace{0.2cm}

\begin{center}

\vspace{0.3cm}

{\it
Bogoliubov  Laboratory of Theoretical Physics, JINR,
141980 Dubna, Russia}

\vspace{0.5cm}

{\tt  nkozyrev@theor.jinr.ru, krivonos@theor.jinr.ru}
\end{center}
\vspace{2cm}

\begin{abstract}
\noindent In this paper we demonstrate that the different generalizations of the Schwarzians, supersymmetric or purely bosonic, can be easily constructed  by using the nonlinear realizations technique.
\end{abstract}


\vskip 0.5cm


\newpage

\setcounter{equation}{0}
\section{Introduction}
The Schwarzian derivative \p{SchwDer}, or just the Schwarzian appears in apparently unrelated fields of mathematics: from classical complex analysis to integrable systems \cite{SW1}. In contrast, in the physics
the Schwarzian appears either in the transformation properties of the conformal (supersymmetric) stress-tensor \cite{Zam,schoutens} or arises as the low energy limit of the SYK model \cite{1}.
Therefore, it is not strange that the possible generalizations of the Schwarzian are mainly related to its supersymmetric extensions where the supersymmetric Schwarzian naturally appears in the superconformal transformations of the super-current \cite{N1Schw,N2Schw,schoutens,MU}. However, this generalization quickly stops at $\cN=4$ supersymmetry due to appearance of components with negative conformal dimension in the current superfield $J^{(N)}$ for $\cN>4$. In addition, the recent
construction of the ``flat space'' version of the Schwarzian \cite{FlatSch} raised the question about
existence of the systematic way to build the generalized (bosonic ones or possessing higher, $\cN > 4$ supersymmetries) Schwarzians.

The treatment of the supersymmetric Schwarzians as the anomalous terms in the transformations of the currents superfield $J^{(\cN)}(Z)$ \cite{schoutens} leads to the conclusion that the structure of the (super)Schwarzians is completely defined by the conformal symmetry and, therefore, it should exist a different, probably purely algebraic, way to define the (super)Schwarzians. The main property of the (super)Schwarzians which defines their structure is their invariance with respect to (super)conformal transformations. The suitable way to construct (super)conformal invariants is the method of nonlinear realizations \cite{coset1,coset2} equipped with the inverse Higgs phenomenon \cite{ih}. However, the natural invariant objects in the non-linear realization approach are the Cartan forms, which contain the differentials of the coordinates of the (super)space with non-trivial transformation properties. Thus, additional question concerns implementation of the inert (super)coordinates in the nonlinear realization approach.

This method was first applied to the $sl(2)$ algebra in \cite{AG1} to obtain the standard Schwarzian and then extended to different super-conformal algebras in \cite{AG2, AG3, AG4, AG5}. Later on, this approach has been applied to the cases of non-relativistic Schwarzians and Carroll algebra \cite{gomis}. It should be noted that  the constraints, proposed in these papers, looks like the results of illuminating guess. Moreover, in some practically interesting cases the proposed constraints are too strong to set the recovered supersymmetric Schwarzian a constant.

In two of our papers \cite{KK1,KK2} the method proposed in \cite{AG1} was modified in two directions.
Firstly, we introduced the ``inert super-space'' as the coordinates of the independent ``inert'' coset elements. Secondly, the constraints were imposed on the full Cartan forms by either nullifying them or identifying with the ``inert super-space'' forms. This last feature gives us the possibility to invoke into game the powerful method of the Maurer-Cartan equations to analyze the consequences of the constraints, which drastically simplified calculations.

In this paper, we apply the proposed approach to construct some new generalized Schwarzians.
After short review of the basic steps of our approach in the Section 2, we will construct
\begin{itemize}
	\item ``Flat space'' variant of the Schwarzian (Section 3),
	\item Bosonic variant of the Schwarzian with $su(1,2)$ symmetry (Section 4),
	\item Schwarzians with $\cN$ extended supersymmetry (Section 5).
\end{itemize}
We conclude in the Section 6 with some interesting but unsolved at the moment questions and hypotheses.

\setcounter{equation}{0}
\section{Sketch of the idea}
Before getting to the main results of our paper, let us illustrate how the method of nonlinear realizations works applied to the Schwarzians on two simpler examples: the standard bosonic Schwarzian ($\cN{=}0$ case) and one of supersymmetric Schwarzians ($\cN{=}2$ case).

\subsection{$\cN{=}0$ case }

The Schwarzian derivative $\left\{ t, \tau\right\}$ is defined by the relation
\be\label{SchwDer}
\left\{ t, \tau\right\} = \frac{\dddot t}{\dot t} - \frac{3}{2} \left( \frac{\ddot t}{\dot t}\right)^2 , \;\; \dot f = \partial_\tau f .
\ee
Its famous property is the invariance with respect to $SL(2,R)$ M\"{o}bius transformations, acting on $t$:
\be\label{sl2a}
t^\prime = \frac{a\, t +b}{c\, t + d} \;\; \Rightarrow \;\; \left\{ t^\prime, \tau\right\} = \left\{ t, \tau\right\}.
\ee
Note, the ``time'' $\tau$ is invariant with respect to these  $SL(2,R)$ transformations.

The action of the bosonic Schwarzian mechanics (see e.g. \cite{1})
\be\label{action0}
S_{schw}[t] = - \frac{1}{2} \int d \tau \left\{ t, \tau\right\}
\ee
leads to the following equation of motion
\be\label{eomschw}
\frac{d}{d\tau}  \left\{ t, \tau\right\} =0 \;\; \Rightarrow \;\;  \left\{ t, \tau\right\} =2\, m^2 = \mbox{const}.
\ee

As $SL(2,R)$ transformations of $t$ are involved, it is natural to look at this system from the nonlinear realization viewpoint. Indeed, one can consider the $sl(2,R)$ algebra, spanned by the Hermitian generators $P, D, K$
\be\label{sl2}
\im \left[ D, P \right] = P,\quad \im \left[ D, K \right] = -K, \quad
\im \left[ K, P \right] = 2 D
\ee
and parameterize the group element in the following way
\be\label{param0}
g = e^{\im t P } e^{\im z K} e^{\im u D}.
\ee
This parameterization is similar to one used in construction of the conformal mechanics \cite{cm}, when $P$, $D$ and $K$ generate time translations, dilatations and conformal boosts, respectively.
Then the Cartan forms, invariant with respect to left multiplication $g^\prime = g_0 g$, read
\be\label{CF0}
g^{-1} d g = \im \omega_P P + \im \omega_D D +\im \omega_K K\;\; \Rightarrow \;\; \omega_P = e^{-u} dt, \quad \omega_D = du - 2 z dt, \quad \omega_K= e^u\left( dz + z^2 dt\right).
\ee
The infinitesimal $sl(2,R)$ transformations
\be\label{sl2tr}
g_0 =  e^{\im \big( \tilde a P+ \tilde b D+ \tilde c K\big)} \;\; \Rightarrow \;\; \delta t = \tilde a + \tilde b t + \tilde c t^2, \;\; \quad \delta u= \frac{d}{dt} \delta t, \quad \delta z = \frac{1}{2}  \frac{d}{dt} \delta u - \frac{d}{dt} \delta t\, z
\ee
are just ones expected for $t$ \p{sl2a}.

If one continues this way, treating $t$ as time and $u$ and $z$ as functions of $t$, imposing covariant condition $\omega_D =0$ would result in elimination of $z$ as an independent variable, $z = \frac{1}{2}\frac{du}{dt}$. (This is manifestation of the Inverse Higgs phenomenon \cite{ih}). Then one can obtain the action of conformal mechanics as \cite{cm}
\be\label{confmechact}
S_{cf} = - \int \left(\omega_K +  m^2 \omega_P\right) = \int dt \left[\frac{1}{4}e^{u}\left(\frac{du}{dt}\right)^2 - m^2 e^{-u}  \right] = \int dt \left[\left(\frac{dx}{dt}\right)^2 -  \frac{m^2}{x^2} \right], \;\; x=e^{u/2}.
\ee
Note, the equation of motion which follows from the action \p{confmechact}
\be\label{eom1}
- e^u \left[ \frac{1}{2}\,\frac{d^2}{dt^2}u + \frac{1}{4}\, \left(\frac{d u}{dt} \right)^2 \right]+m^2 e^{-u}=0
\ee
can be rewritten as the constraint on the Cartan forms \cite{cm}
\be\label{eom2}
\omega_K - m^2 \omega_P = 0.
\ee
Thus the Schwarzian mechanics is essentially the conformal mechanics rewritten in the new coordinates.

The Cartan forms in \p{CF0} are invariant with respect to $sl(2,R)$ transformations \p{sl2tr}, while the time variable $t$ transforms according to \p{sl2a}, \p{sl2tr}. At this point one may impose our main condition \cite{AG1, AG2, KK1}
\be\label{tau}
\omega_P = e^{-u} dt = d\tau,
\ee
where $\tau$ is a new invariant ``time'' which is completely inert under $sl(2,R)$ transformations. Treating now  $t$, $u$, $z$ as the functions of $\tau$, one can express the Goldstone fields $u$ and $z$ in terms of $\dot t$, $\ddot t$
\be\label{uztau}
u= \log{\dot t}, \;\; \omega_D = du - 2\, z\, e^u\, d\tau = 0\; \Rightarrow \; z =\frac{1}{2} e^{-u} \dot{u} = \frac{\ddot t}{2\, {\dot t}{}^2}.
\ee
Putting these relations into the remaining form $\omega_K$, one immediately obtains that it is proportional to the Schwarzian $\big\{ t,\tau\big\}$:
\be\label{omegaK}
\omega_K = \frac{1}{2} d\tau\left[ \ddot u - \frac{1}{2} {\dot u}{}^2 \right]  =
\frac{1}{2} \left[ \frac{\dddot t}{\dot{t}} -\frac{3}{2} \left( \frac{\ddot t}{\dot t}\right)^2 \right] d\tau = \frac{1}{2}d\tau \, \big\{ t,\tau \big\}.
\ee
The Schwarzian action is, obviously, $S_{schw} = - \int \omega_K$.

\subsection{$\cN{=}2$ case }
This idea can be straightforwardly generalized to the supersymmetric case. To obtain supersymmetric Schwarzians, we should consider the proper superalgebra, which differs from \p{sl2} by the presence of supercharges $Q^i$, superconformal charges $S^i$ and, possibly, internal symmetry generators $J^{ij}$. Then one should introduce the superconformally inert superspace coordinates $\tau$ and $\theta$ using the relations
\be\label{susyidea}
\omega_P = \triangle \tau, \;\; \big( \omega_Q\big)^i = d\theta^i,
\ee
where the forms $\triangle \tau$ and $d\theta^i$ are invariant with respect to standard superspace transformations $\delta\tau \sim \epsilon \theta$, $\delta\theta \sim \epsilon$. After imposing condition $\omega_D=0$ also, one should obtain that the remaining forms are composed of supersymmetric Schwarzians and their derivatives. As one of the simplest examples, let us consider $\cN=2$ Schwarzian mechanics \cite{AG2, KK1}.

In the case of $\cN{=}2$ supersymmetry we start from $\cN{=}2$ superconformal algebra $su(1,1|1)$ with the following (anti)commutation relations
\bea\label{N2sca}
&& \im \left[ D, P \right]= P, \quad  \im \left[ D, K \right]= - K, \quad  \im \left[ K, P\right]=2 D,
\nn \\
&& \left\{ Q, \bQ \right\}=2 P, \quad \left\{ S, \bS \right\}=2 K, \quad
\left\{ Q, \bS \right\}=  - 2 D + 2 J, \left\{ \bQ, S \right\}= - 2 D - 2 J,
\nn \\
&& \im \left[ J ,Q \right] =\frac{1}{2} Q, \; \im \left[ J ,\bQ \right] = -\frac{1}{2} \bQ, \quad
\im \left[ J ,S \right] =\frac{1}{2} S, \; \im \left[ J ,\bS \right] = -\frac{1}{2} \bS, \nn \\
&& \im \left[ D, Q \right] = \frac{1}{2} Q,\; \im \left[ D, \bQ \right] = \frac{1}{2} \bQ,\quad
\im \left[D, S \right] = -\frac{1}{2} S,\;\im \left[D, \bS \right] = -\frac{1}{2} \bS, \nn \\
&& \im \left[ K,Q \right] = - S, \; \im \left[ K,\bQ \right] = - \bS, \quad
\im \left[P, S \right]=  Q, \;\im \left[P, \bS \right]=  \bQ .
\eea
We parameterize the $SU(1,1|1)$ group element in the following way
\be\label{N2g}
g=e^{\im t P}\, e^{\xi Q +\bxi \bQ}\, e^{ \psi S+\bpsi \bS} e^{\im z K} e^{\im u D} e^{ \phi J}.
\ee
The Cartan forms
\be\label{N2cfdef}
g^{-1}d g = \im \omega_P P+  \omega_Q Q + {\bar\omega}_Q \bQ + \im \omega_D D+  \omega_J J + \omega_S S + {\bar\omega}_S \bS +\im \omega_K K
\ee
explicitly read
\bea\label{N2cf}
&& \omega_P \equiv e^{-u} \triangle t = e^{-u}\left( dt+ \im (d\bxi \xi  +d\xi \bxi ) \right), \nn \\
&& \omega_Q = e^{-\frac{u}{2}+\im \frac{\phi}{2}}  \left( d\xi +\psi\, \triangle t\right),\;
\bar{\omega}_Q = e^{-\frac{u}{2}-\im \frac{\phi}{2}} \left( d\bxi +\bpsi \,\triangle t\right),\nn \\
&& \omega_D = d u - 2 z\, \triangle t - 2 \im (d\xi\bpsi+d\bxi \psi) , \;
\omega_J = d \phi - 2 \psi \bpsi\, \triangle t +  2 (d\bxi\psi-d\xi \bpsi), \nn \\
&& \omega_S = e^{\frac{u}{2}+\im \frac{\phi}{2}} \left(d \psi -\im \psi \bpsi d\xi + z \left( d\xi +\psi\, \triangle t\right)\, \right), \nn \\
&&\bar{\omega}_S  =  e^{\frac{u}{2}-\im \frac{\phi}{2}} \left(d \bpsi +\im \psi \bpsi d\bxi + z \left( d\bxi +\bpsi \,\triangle t\right)\right),\nn \\
&& \omega_K = e^{u}\left( d z +z^2 \triangle t -\im (\psi\, d \bpsi +\bpsi\, d\psi) +2 \im z\, (d\xi\, \bpsi + d\bxi \psi)\right).
\eea
Now we impose the following conditions on the forms $\omega_P$, $\omega_Q$, $\omega_D$ \p{N2cf}:
\be\label{N2conds}
\omega_P = \triangle \tau, \;\; \omega_Q = d\theta, \;\; {\bar\omega}_Q = d\bar\theta, \;\; \omega_D =0.
\ee
Here, $\triangle \tau = d\tau + \im \big(d\theta \bar\theta + d\bar\theta\theta \big)$. The forms $\triangle\tau$, $d\theta$, $d\bar\theta$ are invariant with respect to $\cN{=}2$ supersymmetry transformations
\be\label{N2tauthetatr}
\delta \tau = \im \big( \epsilon \bar\theta + \bar\epsilon \theta  \big), \;\; \delta\theta = \epsilon, \;\; \delta\bar\theta = \bar\epsilon.
\ee
Covariant derivatives with respect to $\tau$, $\theta$, $\bar\theta$ are
\be\label{N2der}
D = \frac{\partial}{\partial \theta} -i \bar\theta \frac{\partial}{\partial \tau},\;
\bD = \frac{\partial}{\partial {\overline \theta}} -i \theta \frac{\partial}{\partial \tau},  \quad \quad \left\{ D, \bD\right\} = - 2 i \partial_\tau.
\ee
The constraints on the Cartan forms \p{N2conds}, expanded in projections $\triangle\tau$, $d\theta$, $d\bar\theta$ with help of \p{N2der}
\bea\label{N2conds2}
&& \dot{t}+ \im \left( \dot{\bxi}\xi+\dot{\xi}\bxi\right) = e^u, \nn \\
&& \dot{\xi}+ e^u \psi =0, \; D \xi =  e^{\frac{1}{2}(u-\im\, \phi)}, \; \bD \xi =0 , \nn \\
&& \dot{\bxi}+ e^u \bpsi =0, \; \bD \bxi =  e^{\frac{1}{2}(u + \im\, \phi)},\; D \bxi =0 .
\eea
Using \p{N2conds2}, we express all Cartan forms in the terms of $\cN{=}2$ Schwarzian
\be\label{N2Sch}
 \cS_{\cN=2} = \frac{D\dot{\xi}}{D\xi} -\frac{\bD \dot{\bxi}}{\bD\bxi}-2 \im \frac{\dot\xi \dot{\bxi}}{D\xi \bD\bxi}
\ee
as
\bea\label{N2formsfin}
\omega_J = \im  \cS_{\cN=2} \triangle\tau, \;\; \omega_S= -\frac{1}{2} \cS_{\cN=2}\, d\theta-\frac{\im}{2} \bD\cS_{\cN=2} \triangle \tau, \;\;  {\bar\omega}_S = \frac{1}{2} \cS_{\cN=2}\, d{\bar\theta}+\frac{\im}{2}  D \cS_{\cN=2} \triangle \tau,\label{finformsN2} \\
\omega_K =- \frac{1}{2} d\theta  D \cS_{\cN=2} + \frac{1}{2}  d\bar\theta\bD \cS_{\cN=2} +\frac{1}{4}\triangle\tau \left( \im \left[ D, \bD\right] \cS_{\cN=2}- \cS_{\cN=2}^2 \right). \nn
\eea
Note that the same conclusion about the structure of the forms can be achieved by the analysis of Maurer-Cartan equations the forms \p{N2cfdef} satisfy. We will use such equations in Section 5 to study a system with ${\cal N}$ supersymmetries.

The constructed $\cN{=}2$ Schwarzian \p{N2Sch} is invariant with respect to superconformal transformations which explicitly read
\bea\label{N2scftr}
g^\prime = e^{\epsilon Q + \bar\epsilon\bQ}e^{\varepsilon S + \bar\varepsilon \bS} g \quad & \Rightarrow & \quad
\left\{ \begin{array}{l}
	\delta t = \im \big( \bar\epsilon\xi +\epsilon\bxi  \big) - \im t\big( \bar\varepsilon\xi +\varepsilon\bxi  \big), \\
	\delta \xi =\epsilon  - \varepsilon t + \im \varepsilon \xi\bxi, \\
	\delta \bxi = \bar\epsilon - \bar\varepsilon t - \im \bar\varepsilon \xi\bxi.
\end{array} \right.
\eea
Thus one can construct the supersymmetric Schwarzian action as
\be\label{N2schwact}
S_{N2schw} = -\frac{\im}{2} \int d\tau \, d\theta\,d\bar\theta\, \cS_{\cN=2} = -\frac{1}{2}\int \omega_J\, \wedge \,\omega_Q \,\wedge \,{\bar\omega}_Q = \im \int \omega_P\, \wedge\, \omega_S\, \wedge\, {\bar\omega}_Q  =-\im \int \omega_P\, \wedge\, \omega_Q\, \wedge\, {\bar\omega}_S.
\ee
It is matter of a quite lengthy calculation to check that the equations of motion which follow from the action \p{N2schwact} can be written as
\be\label{eom4}
\frac{d}{d\tau}  \cS_{\cN=2} =0 \quad \Rightarrow \quad \cS_{\cN=2} =const = -2\, m .
\ee
Looking at the Cartan forms \p{N2formsfin}, one may note that the equations \p{eom4} reduce them
to the forms on the subalgebra formed by the generators
\be\label{red}
R = P+m^2 K - 2\im m J,\quad \Gamma = Q + \im m S, \quad \overline\Gamma = \bQ-\im m \bS, \quad \big\{ \Gamma, \overline\Gamma \big\}=2R.
\ee
The reduction of the Cartan forms on the algebra $su(1,1|1)$ to the forms on the subalgebra \p{red} is
the key ingredient of the covariant reduction used in \cite{scm} to construct $\cN{=}2$ superconformal mechanics. Thus, in the $\cN=2$ supersymmetric case the Schwarzian mechanics is nothing but the superconformal mechanics written in the superfields $\{t , \xi, \bxi \}$ depending on the coordinates of the inert superspace $\{\tau, \theta, \bar\theta\}$. Unfortunately, this relation does not work beyond
$\cN=2$ case with $m \neq 0$.

\setcounter{equation}0
\section{Flat space analogue of the Schwarzian}
As the first example of the generalized Schwarzian in this Section we will consider the nonlinear realization of the Maxwell algebra which will result in so-called flat space analogue of the Schwarzian. The latter was discovered in \cite{FlatSch} as a result of search of a holographic dual to the Jackiw-Teitelboim gravity \cite{JT1,JT2} in flat space, much like the Sachdev-Ye-Kitaev model which is related to the standard Schwarzian and provides the holographic dual to the JT gravity in AdS space.

The Maxwell algebra contains the Hermitian generators of translation $P$, analogue of the dilatation -- central charge generator $Z$, analogue of the conformal boost $K$, and the generator of $U(1)$ rotations obeying   the following relations
\be\label{maxwell}
\im \left[ J, P \right] = P,\quad \im \left[ J, K \right] = -K, \quad
\im \left[ K, P \right] = 2 Z .
\ee
If we parameterize the Maxwell - group element $g$ as
\be\label{parammaxw}
g = e^{\im t \left( P+q J + m^2 K\right)} e^{\im z K} e^{\im u Z} e^{\im \phi J} ,
\ee
then the Cartan forms
\be\label{CFmaxw}
g^{-1} d g = \im \omega_P P + \im \omega_Z Z +\im \omega_K K +\im \omega_J J
\ee
will read
\be\label{forms0}
\omega_P = e^{-\phi} dt, \quad \omega_Z = du - 2 z dt, \quad \omega_K= e^\phi\left( dz - q z dt +m^2 dt\right),\quad \omega_J = d\phi.
\ee
The constraints
\be
\omega_P = d \tau, \quad \omega_Z = 0
\ee
result in the following relations
\be
\dot{t} = e^\phi,\quad z =\frac{\dot{u}}{2\dot{t}}\; .
\ee
Finally,
\be
\omega_K =d \tau\, \dot{t} \left[ \frac{1}{2} \left( \frac{\ddot{u}}{\dot{t}} - \frac{\dot{u}\ddot{t}}{\dot{t}^2}\right)+ m^2 {\dot t} -\frac{1}{2} q\, \dot{u}\right] \equiv d\tau \cS_{flat}.
\ee
This is exactly the flat space analogue of the Schwarzian constructed in \cite{FlatSch}. Unfortunately, the
simplest invariant action
$$
{\cal S} \sim \int  \omega_K
$$
has no clear geometric interpretation.

\setcounter{equation}{0}
\section{Schwarzian with $su(1,2)$ symmetry  }
In this Section we will consider the bosonic version of the $\cN=2$ superconformal mechanics -- system with $su(1,2)$ symmetry.

The $su(1,2)$ algebra includes the following generators:
\begin{itemize}
	\item the generators $P,D, K$, forming $sl(2,\mathbb{R})$  subalgebra
	\item the generators $Q, \bQ$, and $S,\bS$  - the bosonic analogs of the supersymmetric and conformal supersymmetry generators
	\item $U(1)$ generator $U$
\end{itemize}
The generators $P,D, K$ and $U$ are Hermitian, while the $Q$ and $S$-generators obey the conjugation rules $\left( Q\right)^\dagger =\bQ, \; \left( S\right)^\dagger =\bS $. The non-zero commutators read
\bea\label{osp}
&& \im \left[ P,K\right] = - 2 D,  \; \im \left[ P,D\right] =-P, \;
\im \left[K,D\right]=K,  \nn \\
&& \im \left[ P, S \right] = -Q ,\; \im \left[ P, \bS \right] = -\bQ , \quad
\im \left[ K, Q \right] = S ,\; \im \left[ K, \bQ \right] = \bS , \nn \\
&& \im \left[ D,Q\right] = \frac{1}{2} Q, \;  \im \left[ D,\bQ\right] = \frac{1}{2} \bQ, \quad
\im \left[ D,S\right] = -\frac{1}{2} S, \;  \im \left[ D,\bS\right] = -\frac{1}{2} \bS,  \nn \\
&&  \left[ U, Q\right] = Q,\;  \left[ U, \bQ\right] = -\bQ,\quad
\left[ U, S\right] = S,\;  \left[ U, \bS\right] = -\bS, \nn \\
&& \left[ Q, \bQ  \right]= - \gamma P , \;  \im \left[ Q, \bS \right]= -\frac{3}{2} \gamma U -\im \gamma D, \quad
\left[ S, \bS  \right]= - \gamma K , \;  \im \left[ S, \bQ \right]= \frac{3}{2} \gamma U -\im \gamma D .
\eea
We parameterize the group element in a standard way as
\be\label{g}
g=e^{\im t P} e^{\im \left( \phi Q + \bphi \bQ\right) } e^{\im \left( v S + \bar{v} \bS \right) } e^{\im z K} e^{\im u D} e^{\im \varphi U} .
\ee

The Cartan forms read
\bea
\omega_P &=& e^{-u} \left(  dt +\frac{\im}{2} \gamma \left( \phi  d\bphi - \bphi d \phi\right)\right) \equiv e^{-u} \triangle t,  \nn  \\
\omega_D &=& d u -\im \gamma \left( \bar{v}  d\phi -v d\bphi \right) - 2 z \triangle t  , \nn \\
\omega_K &=& e^u\left[ dz + \left(z^2 +\frac{\gamma^2}{4} v^2 \bar{v}^2\right)\triangle t -  \im \gamma z  \left(v d \bphi -\bv d\phi  \right)+\frac{\im}{2} \gamma \left( v d\bv -\bv dv\right)-\frac{\gamma^2}{2} v \bv \left( v d\bphi +\bv d \phi\right)\right] \nn \\
\omega_Q & = & e^{-\frac{u}{2} -\im\varphi} \left[ d \phi - v \triangle t \right], \quad
{\bar\omega}{}_Q  =  e^{-\frac{u}{2} -\im\varphi} \left[ d \bphi - \bv \triangle t \right], \nn \\
\omega_S & = & e^{\frac{u}{2} -\im \varphi}\left[ d v -
\left( z +\frac{\im}{2} \gamma v \bv\right)\left( d\phi - v \triangle t\right)- \im \gamma v^2 d \bphi \right], \nn \\
{\bar\omega}_S & = & e^{\frac{u}{2} +\im \varphi}\left[ d \bv -
\left( z -\frac{\im}{2} \gamma v \bv\right)\left( d\bphi - \bv \triangle t\right)+ \im \gamma\bv^2 d \phi\right], \nn \\
\omega_U & = & d \varphi -\frac{3}{2} \gamma \left(v d \bphi + \bv d\phi - v \bv \triangle t \right) .
\label{CFgamma}
\eea

The constraints we are going to impose are of three different types:
\begin{itemize}
	\item The constraint which introduce the inert ``time'' $\tau$: ${\omega}_P= d \tau$
	\item The constraints realizing the Inverse Higgs phenomenon \cite{ih}:
	${\omega}_D= {\omega}_Q = {\bar\omega}_Q =0$,
	\item The dynamical constraints which produced the equations of motion:
	${\omega}_K= {\omega}_S = {\bar\omega}_S =0$.
\end{itemize}
As the results of two first constraints we have
\be\label{IH1}
\dot{t} +\frac{\im}{2}\gamma \left( \phi \dot{\bar\phi}-\bar\phi \dot\phi\right) = e^u, \quad v = e^{-u} \dot\phi, \; \bv = e^{-u} \dot\bphi , \quad z= \frac{1}{2} e^{-u} {\dot u}.
\ee
The dynamical constraints give the equations of motion:
\bea\label{eomsu12}
\ddot\phi & = & {\dot u} \dot\phi + \im e^{-u} \gamma \dot\phi{}^2 \dot\bphi, \quad
\ddot\bphi  =  {\dot u} \dot\bphi - \im e^{-u} \gamma \dot\phi \dot\bphi{}^2, \nn \\
{\ddot u} & = & \frac{1}{2} \left( {\dot u}{}^2 - e^{-2 u} \gamma^2 \dot\phi{}^2 \dot\bphi{}^2 \right).
\eea
To get the Schwarzian-like system one has to pass from the variable $u$ to the ``old'' time $t$   \p{IH1}. The result
\be\label{eoumsu121}
\frac{\dddot t + \frac{\im \gamma}{2}  \big( \dot\phi \ddot{\bphi} - \dot\bphi \ddot\phi + \phi \dddot\bphi + \bphi \dddot \phi  \big)}{\dot t + \frac{\im \gamma}{2}\big( \phi \dot{\bphi} - \bphi \dot\phi\big) } - \frac{3}{2} \frac{\left(  \ddot t + \frac{\im \gamma}{2}\big( \phi \ddot{\bphi} - \bphi \ddot\phi\big)  \right)^2}{\left( \dot t + \frac{\im \gamma}{2}\big( \phi \dot{\bphi} - \bphi \dot\phi\big)   \right)^2} = - \frac{1}{2} \frac{\gamma^2 \dot\phi^2 \dot\bphi^2}{\left( \dot t + \frac{\im \gamma}{2}\big( \phi \dot{\bphi} - \bphi \dot\phi\big)   \right)^2}
\ee
is somewhat complicated, but evidently generalizes the equation of motion of the standard Schwarzian mechanics
\be\label{eomschw1}
\frac{\dddot t}{\dot t} - \frac{3}{2}\left( \frac{\ddot t}{\dot t}  \right)^2 = 2m^2.
\ee

\setcounter{equation}{0}
\section{Schwarzians with higher ($\cN > 4$) supersymmetry}
It is well known for a long time that the supersymmetric Schwarzians appear in the transformations
of the current superfield $J^{(N)}(Z)$ under $\cN$-extended superconformal algebra \cite{schoutens}.
In fact, such appearance of the supersymmetric Schwarzians can be considered as their definition.
However, $\cN$ extended superconformal theories have the natural upper bound $\cN=4$, since for $\cN > 4$
the current superfield $J^{(N)}(Z)$  has components with negative conformal dimension.
In the series of the previous papers \cite{KK1,KK2} and related but using a slightly different approach \cite{AG2,AG3},
all such  $\cN=4$ super-Schwarzians were reproduced. As expected, they coincide with the
Schwarzians from the seminal paper by K.~Schoutens \cite{schoutens}.

However, the approach developed in \cite{AG2,AG3,KK1,KK2} and which we advocated here, does not
possess the upper bound on the number of supersymmetries. So, it is natural to try to construct some analogues of the Schwarzian with higher,  $\cN > 4$ supersymmetry. Alas, our first attempts in this direction were failures. The analysis of the superalgebras $osp(4^\star|4)$, and $su(1,1|\cN/2>2)$
leads to conclusion that as result of standard constraints imposed on the differential forms
\be\label{stanconstr}
\omega_P = \triangle \tau, \;\; \big(\omega_Q \big){}^\alpha = d\theta{}^\alpha, \;\;  \big(\bar\omega_Q \big)_\alpha = d\bar\theta_\alpha, \;\; \omega_D =0
\ee
all others, in contrast to the already studied cases with ${\cal N}\leq 4$, are put to zero leaving no room for the Schwarzians in the standard sense. Instead, in such approach we obtain a set of higher-order differential equations on the fields involved, which can be treated as describing some dynamical (and quite possibly integrable) system. Though discussion of these dynamical systems is out of scope of this paper, let us note that there exists at least one possibility when the standard constraints are not strong enough to put the Schwarzian to zero for ${\cal N}\geq 4$. It is given by the series of $osp(\cN|2)$ superalgebras, which we discuss in detail.

\subsection{Superalgebra $osp(\cN|2)$}
The bosonic part of the superalgebra $osp(\cN|2)$ contains among the subgroups $sl(2)\times so(\cN)$
with the generators $(P,D,K)$ and $J_{ij} =-J_{ji},\; i,j = 1,2,\ldots , \cN$, respectively \cite{Sorba}.
The fermionic part of this algebra includes $2\cdot \cN$ fermionic generators $Q_i, S_i$ forming the vectors with respect to $so(\cN)$ algebra and doublet with respect to $sl(2)$ subalgebra. The commutation relations have rather compact form:
\bea\label{osp6vectalg}
&&\big[ D,P  \big]=-\im P, \;\; \big[ D,K  \big]=\im K, \;\; \big[ P,K  \big]=2\im D,  \nn\\
&&\big[ D, Q_i \big] = -\sfrac{\im}{2}Q_i,\;\; \big[ D, S_i\big] = \sfrac{\im}{2}S_i,\;\; \big[ K, Q_i \big] = \im S_i,\;\; \big[ P, S_i\big] = -\im Q_i \nn \\
&&\big\{ Q_i ,Q_j   \big\}=2\delta_{ij} P, \;\; \big\{ S_i ,S_j   \big\}=2\delta_{ij} K, \quad
\big\{ Q_i ,S_j   \big\}=-2\delta_{ij} D + J_{ij}, \nn \\
&&\big[ J_{ij}, J_{kl}  \big] = \im \big( \delta_{ik} J_{jl}  - \delta_{jk} J_{il} -  \delta_{il} J_{jk}+  \delta_{jl} J_{ik} \big), \nn \\
&&\big[ J_{ij}, Q_k  \big] = \im \big( \delta_{ik} Q_j - \delta_{jk} Q_i \big), \;\; \big[ J_{ij}, S_k  \big] = \im \big( \delta_{ik} S_j - \delta_{jk} S_i \big).
\eea
The group element can be defined as
\be\label{osp6vectcoset}
g = e^{\im t P}e^{\xi_i Q_i} e^{\psi_i S_i}e^{\im z K}e^{\im u D} e^{\lambda_{ij} J_{ij}}.
\ee
Here, the superfields $t$, $\xi_i$, $u$, $z$, $\psi_i$, $\lambda_{ij}$ depend on the coordinates of $\cN$-extended
superspace  $\tau$, $\theta_i$. Defining the Cartan  forms as
\be\label{Cforms}
\Omega = g^{-1}d{g} = \im \omega_P P + \im \omega_K K +\im \omega_D D + \big(\omega_Q\big)_i Q_i + \big(\omega_S\big)_i S_i + \im \big( \omega_J \big)_{ij} J_{ij},
\ee
one may impose the standard constraints of our approach \cite{KK1,KK2}
\be\label{mconstr}
\omega_P = \triangle\tau \equiv d\tau + \im d\theta_i \theta_i,\quad \big( \omega_Q  \big)_i = d\theta_i, \quad \omega_D =0 .
\ee
Here, $\tau$ and $\theta_i$ are the coordinates of the ``inert'' superspace. The covariant (with respect to the flat $\cN$-extended supersymmetry, generated by $Q_i$ and $P$) differentials $\triangle \tau$ and $d\theta_i$ can be used to define the covariant derivatives as
\be
d {\cal A} = \triangle \tau\, D_\tau {\cal A} + d \theta_i D_i {\cal A},
\ee
with
\be
D_\tau = \partial_\tau, \; D_i = \frac{\partial}{\partial \theta_i} -\im\, \theta_i \partial_\tau \quad\Rightarrow \quad \left\{ D_i, D_j\right\} = - 2 \im\, \delta_{ij}\, \partial_\tau .
\ee

\subsection{Maurer-Cartan equations}
One may explicitly calculate the forms i \p{Cforms} and analyze the consequences of the constraints \p{mconstr}. However, in practice this way is a rather cumbersome and involved. The simplification comes from the evident statement that our constraints include the Cartan forms themselves,
and, therefore, it makes sense to use the Maurer-Cartan equations to analyze their consequences.

If the Cartan form $\Omega$ is defined as in \p{Cforms},
then by construction it obeys the Maurer-Cartan equation\footnote{Here, $d_1$ and $d_2$ are mutually commuting differentials,  $d \tau$ is the commuting bosonic object, while $d\theta$ is anti-commuting fermionic one.}
\be\label{MaurerCartan}
d_2\Omega_1 - d_1 \Omega_2 = \big[ \Omega_1, \Omega_2   \big], \;\; \Omega_1 = \Omega(d_1), \;\; \Omega_2 = \Omega(d_2).
\ee
This equation can be expanded into following set of equations
\bea\label{osp6vectMC}
\im \big( d_2 \omega_{1P} -d_1\omega_{2P} \big) &=& -\im \big( \omega_{1P}\omega_{2D}- \omega_{1D}\omega_{2P} \big)-2\big( \omega_{1Q}\big)_i \big( \omega_{2Q}  \big)_i, \nn \\
\im \big( d_2 \omega_{1K} -d_1\omega_{2K} \big) &=& \im \big( \omega_{1K}\omega_{2D}- \omega_{1D}\omega_{2K} \big)-2\big( \omega_{1S}\big)_i \big( \omega_{2S}  \big)_i,\nn \\
\im \big( d_2 \omega_{1D} -d_1\omega_{2D} \big) &=& -2\im \big( \omega_{1P}\omega_{2K}- \omega_{1K}\omega_{2P} \big)+2\big( \omega_{1Q}\big)_i \big( \omega_{2S}  \big)_i - 2\big( \omega_{2Q}\big)_i \big( \omega_{1S}  \big)_i,  \\
\im \big( d_2 \big(\omega_{1J}\big)_{ij} -d_1 \big(\omega_{2J}\big)_{ij} \big) &=&  2\im  \big( \omega_{1J}  \big)_{ik}\big( \omega_{2J}  \big)_{kj}  -  2\im  \big( \omega_{2J}  \big)_{ik}\big( \omega_{1J}  \big)_{kj} - \nn \\
&&   \big( \omega_{1Q}\big)_{[i}  \big( \omega_{2S}  \big)_{j]}+ \big( \omega_{2Q}\big)_{[i}  \big( \omega_{1S}  \big)_{j]},  \nn \\
d_2 \big(\omega_{1Q}\big)_i  -d_1 \big(\omega_{2Q}\big)_i &=& \omega_{1P}  \big( \omega_{2S} \big)_i -  \omega_{2P}  \big( \omega_{1S} \big)_i + \frac{1}{2}\big( \omega_{1D}  \big( \omega_{2Q} \big)_i  -  \omega_{2D}  \big( \omega_{1Q} \big)_i   \big) +\nn \\
&&+2 \big(  \omega_{1J} \big)_{ik} \big(  \omega_{2Q} \big)_k - 2 \big(  \omega_{2J} \big)_{ik} \big(  \omega_{1Q} \big)_k, \nn \\
d_2 \big(\omega_{1S}\big)_i  -d_1 \big(\omega_{2S}\big)_i &=& -\omega_{1K}  \big( \omega_{2Q} \big)_i +  \omega_{2K}  \big( \omega_{1Q} \big)_i - \frac{1}{2}\big( \omega_{1D}  \big( \omega_{2S} \big)_i  -  \omega_{2D}  \big( \omega_{1S} \big)_i   \big) +\nn \\
&&+2 \big(  \omega_{1J} \big)_{ik} \big(  \omega_{2S} \big)_k - 2 \big(  \omega_{2J} \big)_{ik} \big(  \omega_{1S} \big)_k. \nn
\eea
To analyze the consequences of these constraints  let us represent other forms in most general way as
\be\label{vectforms}
\big( \omega_S   \big)_i = \triangle\tau \Psi_i + d\theta_j A_{ij}, \;\; \big( \omega_J \big)_{ij} = \triangle\tau X_{ij} + \im d\theta_k \Sigma_{kij}, \;\; \omega_K = \triangle\tau C + \im d\theta_i \, \Xi_i,
\ee
Substituting constraints \p{mconstr}  and the Anzatz for other forms \p{vectforms} into equations \p{osp6vectMC}, one finds that $d\omega_P$ equation is satisfied identically, and $d\omega_Q$ equation implies that
\be\label{osp6vectdomegaQ}
A_{ij} +2 X_{ij}=0, \;\; \Sigma_{kil}+ \Sigma_{lik}=0.
\ee
As by definition $\Sigma_{kij} = -\Sigma_{kji}$, the second equation implies that $\Sigma_{kij}$ is completely antisymmetric. The second, $d\omega_J$ equation reads
\bea\label{osp6vectdomegaJ}
\im D_k X_{ij} + {\dot \Sigma}_{kij} = -2X_{in}\Sigma_{knj} + 2 X_{jn}\Sigma_{kni} + \frac{1}{2}\big( \delta_{ik}\Psi_j - \delta_{jk}\Psi_i    \big), \nn \\
-2\delta_{kl}X_{ij} - \delta_{ik}X_{jl}+\delta_{jk}X_{il} -\delta_{il}X_{jk}+ \delta_{jl}X_{ik} = D_k \Sigma_{lij} + D_l \Sigma_{kij} -2\im \Sigma_{kin}\Sigma_{ljn} -2\im \Sigma_{lin}\Sigma_{kjn}.
\eea
The $d\omega_D$ equation implies that
\be\label{osp6vectdomegaD}
2\Xi_k - 2\Psi_k =0, \;\; A_{ij}+A_{ji}=0.
\ee
The second equation is satisfied due to $A_{ij} = -2X_{ij} = -2X_{[ij]}$. \\
$d\omega_S$ equation reads
\bea\label{osp6vectdomegaS}
D_k \Psi_i - {\dot A}_{ik} = -\delta_{ik} C +2 X_{ij}A_{jk} -2\im \Sigma_{kij}\Psi_j, \nn \\
D_l A_{ik} + D_k A_{il} + 2\im \delta_{kl}\Psi_i = \im \big(  \delta_{il}\Xi_k + \delta_{ik}\Xi_l  \big)-2\im \Sigma_{kij}A_{jl} -2\im \Sigma_{lij}A_{jk}.
\eea
Finally, $d\omega_K$ equation reads
\be\label{osp6vectdomegaK}
\im D_k C + {\dot\Xi}_k = 2\Psi_i A_{ik}, \;\; -2\delta_{kl}C - D_k \Xi_l - D_l \Xi_k =-2A_{km}A_{lm}.
\ee
Taking into account simple equations \p{osp6vectdomegaQ}, \p{osp6vectdomegaD}, one may note that the second equation \p{osp6vectdomegaS} is a direct consequence of the first in \p{osp6vectdomegaJ}, and the second equation \p{osp6vectdomegaK} follows from the first ones in \p{osp6vectdomegaS} and \p{osp6vectdomegaQ}. Therefore, the really independent variables are $\Sigma_{ijk}= \Sigma_{[ijk]}$, $X_{ij} = X_{[ij]}$, $\Psi_i$ and $C$. They satisfy the following set of equations
\bea
-2\delta_{kl}X_{ij} - \delta_{ik}X_{jl}+\delta_{jk}X_{il} -\delta_{il}X_{jk}+ \delta_{jl}X_{ik} =\nn \\= D_k \Sigma_{lij} + D_l \Sigma_{kij} -2\im \Sigma_{kin}\Sigma_{ljn} -2\im \Sigma_{lin}\Sigma_{kjn}, \label{osp6vectall1} \\
\im D_k X_{ij} + {\dot \Sigma}_{kij} = -2X_{in}\Sigma_{knj} + 2 X_{jn}\Sigma_{kni} + \frac{1}{2}\big( \delta_{ik}\Psi_j - \delta_{jk}\Psi_i    \big),  \label{osp6vectall2} \\
D_k \Psi_i +2 {\dot X}_{ik} = -\delta_{ik} C -4 X_{ij}X_{jk} -2\im \Sigma_{kij}\Psi_j,   \label{osp6vectall3}\\
\im D_k C + {\dot\Psi}_k = -4\Psi_i X_{ik}. \label{osp6vectall4}
\eea
The first of these equations \p{osp6vectall1} defines $X_{ij}$ in terms of $\Sigma_{ijk}$ and its derivative. Using this solution and equation \p{osp6vectall1} again, one can find $D_k X_{ij}$ and substitute it to the next equation \p{osp6vectall2}. This reduces \p{osp6vectall2} to terms with $\delta_{ij}$-symbols, which allow to find $\Psi_k$. Continuing down this road, one can simplify \p{osp6vectall3} to find $C$ and check that the last one \p{osp6vectall4} becomes just an identity. Therefore, all the superfields $X_{ij}$, $\Psi_i$, $C$ can be expressed in terms of $\Sigma_{ijk}$ satisfying \p{osp6vectall1}. It is remarkable that this can be done for arbitrary number of supersymmetries $\cN$. The solution explicitly reads
\bea\label{comp}
X_{ij} & = & \frac{1}{2-{\cal N}}\big( D_m \Sigma_{mij} -2\im \Sigma_{imn}\Sigma_{jmn}   \big), \nn \\
\Psi_i & = & - \frac{2\im}{{\cal N}-1}\big( D_l X_{il} +2\im X_{mn}\Sigma_{imn}    \big), \nn \\
C & = & -\frac{1}{{\cal N}} \big( D_j \Psi_j -4 X_{mn}X_{mn}   \big) .
\eea
Thus, all the Cartan forms can be expressed in terms of unique object -- superfield $\Sigma_{ijk}$.
This superfield $\Sigma_{ijk}$, being fully anti-symmetric over permutations of the indices, appeared as $d\theta$-projection of the form $(\omega_J)_{ij}$. Due to these properties, one can call this superfield  $\Sigma_{ijk}$ as the supersymmetric $\cN$ - extended Schwarzian. It is satisfies the nonlinear constraint given by equation \p{osp6vectall1}, where $X_{ij}$ is expressed in terms of $\Sigma_{ijk}$ by \p{comp}.

\subsection{Explicit form of the supersymmetric $\cN$-extended Schwarzian}
From the previous Subsection, we see that the supersymmetric $\cN$ - extended Schwarzian $\Sigma_{ijk}$ we are looking for, appears as $d\theta$ - projection of the form $(\omega_J)_{ij}$. Thus, the final task is to express  $\Sigma_{ijk}$ in terms of the parameters of the group element \p{osp6vectcoset}, depending, in virtue of our constraints \p{mconstr} from the coordinates of the flat inert superspace
$\tau, \theta_i$.

The Cartan forms $\Omega = g^{-1}dg$, explicitly calculated for the group element \p{osp6vectcoset}, read
\bea\label{ospvect62CF}
\omega_P = e^{-u}\triangle t = e^{-u}\big(  dt + \im d\xi_j \, \xi_j \big), \;\; \big( \omega_Q    \big)_i = e^{-u/2} \big( d\xi_j + \triangle t \psi_j   \big) M_{ji}, \nn \\
\omega_D = du -2\im d\xi_k \, \psi_k -2z \triangle t, \;\; \big(\omega_S \big)_i = e^{u/2} \big( d\psi_j + \im d\xi_k\, \psi_k\,\psi_j + z\big(d\xi_j + \triangle t\psi_j\big) \big)M_{ji},\nn \\
\omega_K = e^u\big( dz + z^2 \triangle t + \im d\psi_j \, \psi_j +2\im z d\xi_j\, \psi_j \big), \nn \\
\big(\omega_J \big)_{ij} = \frac{1}{2}\big( M^{-1} \big)_{km}d M_{ml} + \frac{\im}{2} \big( M^{-1} \big)_{km} \big( M^{-1} \big)_{nl}e^{-u}\big( d\xi_m \, \psi_n - d\xi_n \, \psi_m  + \triangle t \psi_m \psi_n  \big).
\eea
Here, the $so(\cN)$ matrix $M_{ij}$ is defined as
\be\label{ospvect6M}
M_{ij} = \big(  e^{2\lambda} \big)_{ij} = \delta_{ij} +2 \lambda_{ij} + \frac{4 \lambda_{ik}\lambda_{kj}}{2!} + \frac{8 \lambda_{ik}\lambda_{kl}\lambda_{lj}}{3!} + \ldots, \;\; \big( M^{-1} \big)_{ij} = M_{ji}.
\ee

The constraints \p{mconstr} imply
\bea\label{ospvect6constr}
\dot t + \im \dot\xi_k \, \xi_k = e^{u}, \;\; D_i t + \im D_i \xi_k \, \xi_k =0, \nn \\
D_m \xi_k = e^{u/2} \big( M^{-1}  \big)_{mk}, \;\; \psi_i = -e^{-u}{\dot\xi}_i, \nn \\
z = \frac{1}{2}e^{-u} \dot u, \;\; D_i u = 2\im D_i \xi_j \,\psi_j.
\eea
Some of these relations define some of the Goldstone fields in terms of derivatives of others, some are not independent. For example, acting by $D_j$ on the second relation and symmetrizing w.r.t. $i,j$ one can obtain
\be\label{ospvect6cons1}
D_i \big( D_j t + \im D_j \xi_k \, \xi_k \big)  + D_j \big( D_i t + \im D_i \xi_k \, \xi_k \big) =0 \; \Rightarrow \; \big( \dot t + \im \dot\xi_k \, \xi_k  \big)\delta_{ij} = D_i\xi_k \, D_j \xi_k.
\ee
Therefore, $D_i\xi_k$ has structure implied by the third equation \p{ospvect6constr}. From this, it can also be derived that
\bea\label{ospvect6cons2}
D_i e^u = \frac{1}{\cN} D_i \big( D_m \xi_n\,  D_m \xi_n   \big) = \frac{2}{\cN} D_i D_m \xi_n\,  D_m \xi_n   = \frac{2}{\cN} \big( -2\im D_i \xi_n {\dot\xi}_n - D_m D_i \xi_n \, D_m \xi_n  \big) = \nn \\
= \frac{2}{\cN} \big( -2\im D_i \xi_n {\dot\xi}_n - \im \cN D_i \xi_n {\dot\xi}_n - D_i e^u   \big) \;\; \Rightarrow D_i e^u = -2\im D_i \xi_m {\dot \xi}_m.
\eea
Taking into account all known kinematic equations, the $d\theta$-projection of the form  $\omega_J$ reads
\be\label{ospvect6cons3}
\big(\omega_J \big)_{kl} = \ldots + d\theta_p \big[ \frac{1}{2}  \big(  M^{-1} \big)_{kn} D_p M_{nl} -\frac{\im}{2}\big( D_p \xi_m {\dot\xi}_n  - D_p \xi_n {\dot\xi}_m \big)e^{-u} \big(M^{-1} \big)_{km} \big(M^{-1} \big)_{ln}  \big] \equiv \ldots + \im d\theta_p \Sigma_{pkl}.
\ee
This expression can be further simplify leading to the following supersymmetric Schwarzian:
\be\label{ospvect6schwfin}
\Sigma_{pkl} = \im\,  \frac{1}{2}e^{-u} D_{[p}D_k \xi_m \, D_{l]}\xi_m = \frac{\im \cN}{2} \frac{D_{[i}D_j \zeta_m \, D_{k]}\zeta_m}{ D_p \zeta_q D_p \zeta_q}.
\ee
The standard bosonic Schwarzian is hidden inside the components of the third derivative of $\Sigma_{ijk}$.  Roughly speaking, the bosonic part of the Schwarzian reads
\be\label{ospvect6DDDSigma}
\frac{2\im D_m D_n D_p \Sigma_{mnp} }{\cN(\cN-1)(\cN-2)}  \approx -\frac{1}{2} \Big( \frac{\dddot t}{\dot t} - \frac{3}{2}\frac{\ddot t {}^2}{\dot t {}^2}   \Big)+ 4 \frac{D_{[k}\Sigma_{lij]}D_{[k}\Sigma_{lij]}}{\cN(\cN-1)(\cN-2)} + 3 \frac{\cN-2}{\cN(\cN-1)}{\dot M}_{mn}{\dot M}_{mn}.
\ee
Note that $D_{[k}\Sigma_{lij]}$ is absent in equation \p{osp6vectall1} and starts from an independent component.

As we are discussing arbitrarily high supersymmetries, it is natural to ask whether the main constraint $D_i t + \im D_i \xi_j \, \xi_j =0$ puts the system on shell. Explicit component analysis of this constraint for some values of $\cN$ indicates, however, that it is essentially an algebraic one, defining superfields $t$ and $\xi_i$ in terms of some unconstrained scalar superfield.

\subsection{Properties with respect to coordinate changes}
The supersymmetric Schwarzian should possess a special property with respect to coordinate changes, known as the composition law. The coordinate changes are diffeomorphism transformations
\be\label{diffeomorph}
\theta_i \rightarrow \theta^\prime_i = \tilde\theta_i (\tau,\theta), \;\; \tau^\prime = \tilde\tau(\tau, \theta),
\ee
constrained by $D_i \tilde \tau + \im D_i\tilde\theta_j \, \tilde\theta_j =0$, so that the derivative $D_i$ transforms homogeneously, $D_i = D_i\tilde\theta_j D^\prime_j$. If the composition law holds for the Schwarzian, it should have the form
\be\label{complaw}
\Sigma_{ijk}\big[ \zeta(\tau^\prime, \theta^\prime); \tau, \theta  \big] = \Sigma_{ijk}\big[ \tilde\theta(\tau^\prime, \theta^\prime);
\tau, \theta  \big] + M_{[ijk]}{}^{mnp} \Sigma_{mnp}\big[ \zeta(\tau^\prime, \theta^\prime); \tau^\prime, \theta^\prime  \big]
\ee
with some matrix $M_{[ijk]}{}^{mnp}$. As the Schwarzian reads
\be\label{NSchwarzian}
\Sigma_{ijk} = \frac{\im \cN}{2} \frac{D_{[i}D_j \zeta_m \, D_{k]}\zeta_m}{ D_p \zeta_q D_p \zeta_q}.
\ee
As for $D_i \zeta_j$ and $D_i \tilde\theta_j$ the relations $D_i \zeta_k D_j \zeta_k \sim \delta_{ij}$ and $D_i \tilde\theta_k D_j \tilde\theta_k \sim \delta_{ij}$ hold, one can shortly obtain
\be
D_p \zeta_q D_p \zeta_q = \frac{D_k \tilde\theta_l D_k \tilde\theta_l}{\cN} D^\prime_p \zeta_q D^\prime_p \zeta_q.
\ee
Then, directly substituting $D_i\zeta_k = D_i\tilde\theta_j D^\prime_j \zeta_k$ into \p{NSchwarzian}, we obtain
\be\label{NSchwarziantr}
\Sigma_{ijk}\big[ \zeta(\tau^\prime, \theta^\prime); \tau, \theta  \big] = \Sigma_{ijk}\big[ \tilde\theta(\tau^\prime, \theta^\prime);
\tau, \theta  \big] + \frac{\cN}{D_r \tilde\theta_s D_r \tilde\theta_s }D_i \tilde\theta_m D_j \tilde\theta_n D_k \tilde\theta_p \Sigma_{mnp}\big[ \zeta(\tau^\prime, \theta^\prime); \tau^\prime, \theta^\prime  \big].
\ee
Thus the Schwarzian transforms as in \p{complaw}, as it should be.

\setcounter{equation}{0}
\section{Conclusion}
In this paper we applied the method of nonlinear realization to some bosonic  (Maxwell algebra and $su(1,2)$ one) and supersymmetric $osp(\cN|2)$ algebras. After introducing the coordinates of the inert (super)spacetime and imposing the proper constraints
$$ \mbox{Cartan forms} = \mbox{Cartan forms on the flat superspace}$$
we expressed all the Cartan forms of the initial (super)algebra through a single object - generalized
Schwarzian. While doing so, we were able to construct the Schwarzians with $\cN$-extended supersymmetry.

The obtained results have to be treated as the first steps in the complete analysis of the Schwarzian systems. Two immediate, but still unanswered questions concern
\begin{itemize}
	\item existence of other $\cN$-extended systems for $\cN > 4$, such as related to $F(4)$ superalgebra,
	\item structure of the equations of motion in Schwarzian supersymmetric mechanics.
\end{itemize}
It is clear that there is no hope to have the superfield actions for the theories with $\cN$ -extended
supersymmetry. However, the question of the equations of motion for such system is not trivial.
As we know, in the bosonic case the equations of motion of the Schwarzian mechanics reduces to the
condition
$$ \mbox{Schwarzian} = const $$
It is interesting to understand whether this property can be extended to the supersymmetric case.
Another interesting continuation concerns supersymmetric Maxwell group, its analysis and possible
relation of the corresponding Schwarzians  with the flat-space analogues of the Sachdev-Ye-Kitaev  model.

Finally, there is a strong expectation that all models constructed in a such manner have to be integrable. It would be interesting to analyze the situation with integrability, at least for the simplest models.

\section*{Acknowledgements}

The work was supported by Russian Foundation for Basic Research, grant
No~20-52-12003.

\end{document}